\documentclass[aps,twocolumn,showpacs,floatfix,amsmath,amssymb]{revtex4}
\usepackage{epsfig}  
\begin{document}

\title{Mass and width of strange baryon resonances in QCD sum rules}
\author{Janardan P. Singh}
\affiliation{Physics Department, Faculty of Science, The M. S. University of Baroda, Vadodara-390002, India}
\author{Frank X. Lee}
\affiliation{Center for Nuclear Studies, Department of Physics,
The George Washington University,  Washington, DC 20052, USA}

\begin{abstract}
The mass spectra of strange baryons in the octet family are investigated 
in a finite0energy QCD sum rule approach based on the Gauss-Weierstrass transform. 
The phenomenological form of the spectral function is saturated by the ground state 
and two of the lowest excited states, considered as having opposite parities. 
Treating the ground state parameters as known, the masses, 
widths and couplings to the interpolating fields are determined 
and compared to experiment.
\end{abstract}
\vspace{1cm}
\maketitle

\section{Introduction}
\label{intro}
An important goal of hadronic physics is to understand the baryon spectrum from QCD, 
the underlying theory of the strong interaction. 
In quantum mechanics, stationary states in bound state problems are eigenstates
(eigenfunction and eigenenergy) of an appropriate Hamiltonian. 
Bound states such as light baryonic systems in QCD, on the other hand,
are far more complicated objects. 
Here the excitation energy is sufficient to create several of the constituents.
In hadrons a typical excitation energy (due to QCD) is few hundred MeV.
This is sufficient to create one or more light quark-antiquark pair/pairs.
QCD is intrinsically a strongly interacting many-body theory. 
Its strong coupling and confinement nature makes it extremely difficult to solve
at low energies relevant for hadronic binding. 
One theoretical tool that has enjoyed success in the low-energy regime is the
QCD sum rule method~\cite{SVZ79}.
It is a non-perturbative approach to QCD that reveals a direct connection between
hadronic observables and the QCD vacuum structure via a few universal parameters
called vacuum condensates (vacuum expectation values of QCD local operators).
It is based on evaluation of a suitable correlation 
function in deep Euclidean region using operator product expansion (OPE) on the one hand, and 
its phenomenological evaluation using physical hadronic states on the other hand.
The method is analytical, physically transparent, and
has minimal model dependence with well-understood
limitations inherent in the OPE.
It provides a complementary view of the same non-perturbative physics 
to the numerical approach of lattice QCD.
The method was applied to the baryon sector not long after it was 
introduced~\cite{Bely81,Bely82,Esp83,RRY82,RRY85,Chung84}, and 
later improved and extended to include some excited 
states~\cite{Dosch89,Derek90,Derek96,Jido97,Lee98,Lee02,Lee06} 
and magnetic moments~\cite{Pasupathy86,Chiu87,Lee98b,Lee98c}.
On the experimental side, the effort is fueled by data of increasing quality 
from JLab and other accelerator facilities. 

In all of the studies, however, the conventional Borel transform and 
pole-plus-continuum ansatz has been used in which the excited states are exponentially suppressed. 
In order to gain access beyond the ground-state pole, one has to seek alternative ansatz in the 
phenomenological spectral function.
Such an attempt has been made in~\cite{Singh94} where the first two 
excited states of the nucleon are explicitly included, 
using a combined framework of Gaussian sum rules and finite energy sum rules (FESR).
As advocated in~\cite{Singh94} and in~\cite{Bert85}, FESR 
is more suitable for studying the spectrum of excited states because its spectral function 
has a polynomial kernel instead of an exponential one.
Furthermore, the widths of the resonances can be taken into account 
in addition to the masses, a unique feature of this procedure. 
In this work, we will extend the work in~\cite{Singh94} to the other members of the baryon octet, 
using an updated version of the Borel sum rules for the octet baryons as given in Ref.~\cite{Lee02}. 
In effect, it is an exercise similar as calculation of excitation energies 
of atoms and nuclei, but using fully relativistic quantum field theoretic tools.
In our calculation, we consistently include operators up to dimension 8 with radiative corrections,
first order strange quark mass corrections,
flavor symmetry breaking of the strange quark condensates,
anomalous dimension corrections, and possible factorization violation of
the four-quark condensate.

\section{Method}
\label{elem}

Hadron masses can be extracted from the time-ordered two-point correlation function in the QCD vacuum,
\begin{equation}
\Pi(p)=i\int d^4x\; e^{ip\cdot x}\,\langle 0\,|\,
T\{\;\eta(x)\, \bar{\eta}(0)\;\}\,|\,0\rangle. \label{cf2pt-old}
\end{equation}
where $\eta$ is the interpolating field with
the quantum numbers of the baryon under consideration. 
\begin{widetext}
For nucleon we use the following interpolating field
\begin{equation}
\eta^{ N}=-2\,\epsilon^{abc}\left[ (u^{aT}C\gamma_5 d^b)u^c +t\,(u^{aT}Cd^b)\gamma_5 u^c \right].
\end{equation}
Here $C$ is the charge conjugation operator and the superscript
$T$ means transpose. The antisymmetric $\epsilon^{abc}$ and sum over color ensures 
a color-singlet state.
The real parameter $t$ allows the freedom for optimal mixing of the two
operators. The choice advocated by Ioffe and usually used in QCD
sum rules studies corresponds to $t=-1$. We will explore different values.
For the other members of the octet family, we consider
for $\Sigma$
\begin{equation}
\eta^{ \Sigma}=-2\,\epsilon^{abc}\left[
(u^{aT}C\gamma_5 s^b)u^c
+t\,(u^{aT}Cs^b)\gamma_5 u^c \right],
\end{equation}
for $\Xi$
\begin{equation}
\eta^{ \Xi}=-2\,\epsilon^{abc}\left[
(s^{aT}C\gamma_5 u^b)s^c
+t\, (s^{aT}Cu^b)\gamma_5 s^c \right].
\end{equation}
For the $\Lambda$, there is the possibility of octet and flavor-singlet quantum numbers.
In this work we consider only the octet member whose 
interpolating field is 
\begin{eqnarray}
\eta^{ \Lambda_o}=
2 \sqrt{1\over6}\, \epsilon^{abc}&& \left\{\left[
 2(u^{aT}C\gamma_5 d^b) s^c
+ (u^{aT}C\gamma_5 s^b) d^c
- (d^{aT}C\gamma_5 s^b) u^c
\right] \right. \nonumber \\& &  \left.
+ t \left[
 2(u^{aT}C d^b)\gamma_5 s^c
+ (u^{aT}C s^b)\gamma_5 d^c
- (d^{aT}C s^b)\gamma_5 u^c
\right] \right\}.
\end{eqnarray}

The most general structure of $\Pi(p)$  is
\begin{equation}
\Pi(p)=\hat{p} F_1(p^2) + F_2(p^2).
\end{equation}
where $\hat{p}\equiv \gamma\cdot p$.
Wilson's operator product expansion up to dimension eight gives the result 
in the following form for the invariant functions $F_1$ and $F_2$
\begin{eqnarray}
-F_1(p^2) &=&
[A+B\ln(-p^2/\mu^2)]\ln(-p^2/\mu^2)] p^4
+ A_4 \ln(-p^2/\mu^2) \langle {\alpha_s\over \pi} G^2\rangle
+ E_4 \ln(-p^2/\mu^2) m_s \langle \bar{q}q\rangle
 \nonumber \\& & 
+ [A_6+B_6 \ln(-p^2/\mu^2)] (1/p^2) \kappa_v \langle \bar{q}q\rangle^2 
+ E_6 (1/p^2) m_s \langle\bar{q}g\sigma\cdot Gq\rangle
+ A_8 (1/p^4) \langle \bar{q}q\rangle \langle\bar{q}g_s\sigma\cdot Gq\rangle 
+ \cdots
\label{F1}
\end{eqnarray}
\begin{eqnarray}
-F_2(p^2) &=&
H_1 \ln(-p^2/\mu^2)] p^4 m_s
+ C_3 p^2 \ln(-p^2/\mu^2) \langle \bar{q}q\rangle
+ [C_5+D_5 \ln(-p^2/\mu^2)] \ln(-p^2/\mu^2) \langle\bar{q}g\sigma\cdot Gq\rangle
 \nonumber \\& & 
+ H_5  \ln(-p^2/\mu^2) m_s \langle {\alpha_s\over \pi} G^2\rangle
+ C_7  (1/p^2)  \langle \bar{q}q\rangle \langle {\alpha_s\over \pi} G^2\rangle
+ H_7 m_s \kappa_v \langle \bar{q}q\rangle^2 
+ \cdots
\label{F2}
\end{eqnarray}
The ellipses denote higher-order terms that are ignored.
Note that $F_1$ contains only dimension-even condensates, 
while $F_2$ contains only dimension-odd condensates.
The coefficients in Eq.~(\ref{F1}) and Eq.~(\ref{F2}) are 
given in the Appendix.

To derive the finite-energy sum rules (FESR), we 
employ the Gauss-Weierstrass (GW) transform of $F_1$ and $F_2$, as outlined in 
Refs.~\cite{Singh94,Bert85}. We compute the lowest three Hermite moments for $F_1$ and $F_2$,
yielding six equations from which six sum rules can be constructed.
Using the abbreviation $(5+2t+5t^2)/[32(2\pi)^4]=\beta$,
the moments, 
after renormalization group improvement and the introduction of a cutoff $s_0$ to 
indicate the onset of the QCD continuum,
are as follows,
\begin{equation}
{1\over\pi}\int^{s_0}_0 \hspace{-1mm} ds\, \mbox{Im} F_1(s) =
  \beta (1+{75\over 12}{\bar{\alpha}_s\over \pi}) L^{-4\over 9} {s_0^3\over 3} 
+ A_4 L^{-4\over 9} \langle {\alpha_s\over \pi} G^2\rangle s_0
+ E_4 L^{-4\over 9} m_s \langle \bar{q}q\rangle s_0
+ \bar{A}_6 \kappa_v \langle \bar{q}q\rangle^2 
+ E_6 L^{-26\over 27} m_s \langle\bar{q}g\sigma\cdot Gq\rangle,
\label{F1a}
\end{equation}
\begin{equation}
{1\over\pi}\int^{s_0}_0 \hspace{-1mm} ds \,s\, \mbox{Im} F_1(s) =
  \beta (1+{25\over 3}{\bar{\alpha}_s\over \pi}) L^{-4\over 9} {s_0^4\over 4} 
+ A_4 L^{-4\over 9} \langle {\alpha_s\over \pi} G^2\rangle {s_0^2\over 2}
+ E_4 L^{-4\over 9} m_s \langle \bar{q}q\rangle {s_0^2\over 2}
+ A_8 \langle \bar{q}q\rangle \langle\bar{q}g\sigma\cdot Gq\rangle,
\label{F1b}
\end{equation}
\begin{equation}
{1\over\pi}\int^{s_0}_0 \hspace{-1mm} ds \,s^2\, \mbox{Im} F_1(s) =
  \beta (1+{367\over 60}{\bar{\alpha}_s\over \pi}) L^{-4\over 9} {s_0^5\over 5} 
+ A_4 L^{-4\over 9} \langle {\alpha_s\over \pi} G^2\rangle {s_0^3\over 3}
+ B_6 \kappa_v \langle \bar{q}q\rangle^2 {s_0^2\over 2}
+ E_4 L^{-4\over 9} m_s \langle \bar{q}q\rangle {s_0^3\over 3},
\label{F1c}
\end{equation}
\begin{equation}
{1\over\pi}\int^{s_0}_0 \hspace{-1mm} ds\, \mbox{Im} F_2(s) =
  H_1 m_s L^{-8\over 9} {s_0^3\over 3} 
+ C_3 \langle \bar{q}q\rangle {s_0^2\over 2}
+ \bar{C}_5 \langle\bar{q}g\sigma\cdot Gq\rangle s_0
+ H_5 L^{-8\over 9} m_s \langle {\alpha_s\over \pi} G^2\rangle {s_0}
+ C_7 \langle \bar{q}q\rangle \langle {\alpha_s\over \pi} G^2\rangle
+ H_7 m_s \kappa_v \langle \bar{q}q\rangle^2,
\label{F2a}
\end{equation}
\begin{equation}
{1\over\pi}\int^{s_0}_0 \hspace{-1mm} ds\,s\, \mbox{Im} F_2(s) =
  H_1 m_s L^{-8\over 9} {s_0^4\over 4} 
+ C_3 \langle \bar{q}q\rangle {s_0^3\over 3}
+ \bar{C}_5 \langle\bar{q}g\sigma\cdot Gq\rangle {s_0^2\over 2}
+ H_5 L^{-8\over 9} m_s \langle {\alpha_s\over \pi} G^2\rangle {s_0^2\over 2},
\label{F2b}
\end{equation}
\begin{equation}
{1\over\pi}\int^{s_0}_0 \hspace{-1mm} ds\,s^2\, \mbox{Im} F_2(s) =
  H_1 m_s L^{-8\over 9} {s_0^5\over 5} 
+ C_3 \langle \bar{q}q\rangle {s_0^4\over 4}
+ \bar{C}_5 \langle\bar{q}g\sigma\cdot Gq\rangle {s_0^3\over 3}
+ H_5 L^{-8\over 9} m_s \langle {\alpha_s\over \pi} G^2\rangle {s_0^3\over 3}.
\label{F2c}
\end{equation}
The above six expressions form the left-hand-side (LHS) of the sum rules.
The two redefined coefficients appearing in the above expressions are given by,
for $\Sigma$,
\begin{equation}
\bar{A}_6= {1\over 6} \left[ (6f_s+1)(1-{43\over 42}{\bar{\alpha}_s\over \pi}) L^{248\over 189}
       - 2 t (1-{1\over 6}{\bar{\alpha}_s\over \pi}) L^{8\over 27}
       - (6f_s-1) t^2 (1-{29\over 30}{\bar{\alpha}_s\over \pi}) L^{184\over 135} \right],
\end{equation}
\begin{equation}
\bar{C}_5= {3\over 4(2\pi)^2} \left[ (1+{50\over 9}{\bar{\alpha}_s\over \pi}) 
       - t^2 (1+{62\over 9}{\bar{\alpha}_s\over \pi}) \right] L^{-14\over 27}.
\end{equation}
For $\Lambda_o$, 
\begin{equation}
\bar{A}_6= {1\over 18} \left[ (10f_s+11)(1-{43\over 42}{\bar{\alpha}_s\over \pi}) L^{248\over 189}
       + (2-8f_s) t (1-{1\over 6}{\bar{\alpha}_s\over \pi}) L^{8\over 27}
       - (2f_s+13) t^2 (1-{29\over 30}{\bar{\alpha}_s\over \pi}) L^{184\over 135} \right],
\end{equation}
\begin{equation}
\bar{C}_5= {1+2f_s\over 4(2\pi)^2} \left[ (1+{179\over 36}{\bar{\alpha}_s\over \pi})
       - t^2 (1+{227\over 36}{\bar{\alpha}_s\over \pi}) \right] L^{-14\over 27}.
\end{equation}
For $\Xi$,
\begin{equation}
\bar{A}_6= {f_s\over 6} \left[ (f_s+6)(1-{43\over 42}{\bar{\alpha}_s\over \pi}) L^{248\over 189}
       - 2f_s t (1-{1\over 6}{\bar{\alpha}_s\over \pi}) L^{8\over 27}
       + (f_s-6) t^2 (1-{29\over 30}{\bar{\alpha}_s\over \pi}) L^{184\over 135} \right],
\end{equation}
\begin{equation}
\bar{C}_5= {3f_s\over 4(2\pi)^2} \left[ (1+{179\over 36}{\bar{\alpha}_s\over \pi})
       - t^2 (1+{227\over 36}{\bar{\alpha}_s\over \pi}) \right] L^{-14\over 27}.
\end{equation}

For the phenomenological side of the spectral function, following Ref.~\cite{Singh94}, 
we take contribution from the lowest three states: 
the ground state with positive parity, the first excited state 
with positive parity, and the second excited state with negative parity.
We treat the ground state as a pole described by two parameters: the coupling strength $\lambda^2$ 
and mass $m$. The two excited states are treated as resonances with finite widths, 
each characterized by three parameters ($\lambda_1^2$, $m_1$, $\Gamma_1$) and 
($\lambda_2^2$, $m_2$, $\Gamma_2$). The spectral functions read
\begin{equation}
{1\over\pi}\mbox{Im} F_1(s) =
  \lambda^2 \delta(s-m^2) 
+ {1\over\pi} \left[ 
  {\lambda^2_1 m_1 \Gamma_1 \over (s-m_1^2+\Gamma_1^2/4)^2 + m_1^2 \Gamma_1^2}
+ {\lambda^2_2 m_2 \Gamma_2 \over (s-m_2^2+\Gamma_2^2/4)^2 + m_2^2 \Gamma_2^2}
\right] \theta(s-m^2),
\end{equation}
\begin{equation}
{1\over\pi}\mbox{Im} F_2(s) =
  \lambda^2 M \delta(s-m^2) 
+ {1\over\pi} \left[ 
  {\lambda^2_1 (s+m_1^2+\Gamma_1^2/4) \Gamma_1/2 \over (s-m_1^2+\Gamma_1^2/4)^2 + m_1^2 \Gamma_1^2}
- {\lambda^2_2 (s+m_2^2+\Gamma_2^2/4) \Gamma_2/2 \over (s-m_2^2+\Gamma_2^2/4)^2 + m_2^2 \Gamma_2^2}
\right] \theta(s-m^2).
\end{equation}
In the limit of zero widths ($\Gamma_i\rightarrow 0$), the excited-state contributions also 
reduce to $\delta$-functions.
The same feature as in the Borel sum rules 
that the excited states with opposite parities add in the chiral-even $F_1$ and 
cancel in the chiral-odd $F_2$ is also preserved.
The $\lambda$'s are the coupling strengths of the interpolating field to the states, defined by 
\begin{equation}
<0|\eta|Bps>=\lambda_B u(p,s),
\end{equation}
where $u(p,s)$ is the spin-1/2 spinor.

Applying the same GW transform, taking the first three Hermite moments of the spectral functions, 
and introducing the cutoff $s_0$, one obtains the following six phenomenological 
expressions~\cite{Singh94} that matches one-to-one to those on QCD side in Eqs~(\ref{F1a}) to (\ref{F2c}):
\begin{equation}
{1\over\pi}\int^{s_0}_0 \hspace{-1mm} ds\, \mbox{Im} F_1(s) =
\lambda^2 + {\lambda^2_1\over\pi} f_1 + {\lambda^2_2\over\pi} f_2,
\label{rhs1a}
\end{equation}
\begin{equation}
{1\over\pi}\int^{s_0}_0 \hspace{-1mm} ds\,s\, \mbox{Im} F_1(s) =
  \lambda^2 m^2 
+ {\lambda^2_1\over\pi} \left[ {m_1\Gamma_1\over 2} r_1 + (m_1^2-\Gamma_1^2/4)f_1 \right]
+ {\lambda^2_2\over\pi} \left[ {m_2\Gamma_2\over 2} r_2 + (m_2^2-\Gamma_2^2/4)f_2 \right],
\label{rhs1b}
\end{equation}
\begin{eqnarray}
{1\over\pi}\int^{s_0}_0 \hspace{-1mm} ds\,s^2\, \mbox{Im} F_1(s) =
  \lambda^2 m^4 
&+& {\lambda^2_1\over\pi} m_1\Gamma_1 \left[ (s_0-m^2) 
+ {(m_1^2-\Gamma_1^2/4)^2 - m_1^2\Gamma_1^2 \over m_1\Gamma_1} f_1 + (m_1^2-\Gamma_1^2/4) r_1 \right]
 \nonumber \\
&+& {\lambda^2_2\over\pi} m_2\Gamma_2 \left[ (s_0-m^2) 
+ {(m_2^2-\Gamma_2^2/4)^2 - m_2^2\Gamma_2^2 \over m_2\Gamma_2} f_2 + (m_2^2-\Gamma_2^2/4) r_2 \right],
\label{rhs1c}
\end{eqnarray}
\begin{equation}
{1\over\pi}\int^{s_0}_0 \hspace{-1mm} ds\, \mbox{Im} F_2(s) =
  \lambda^2 m 
+ {\lambda^2_1\over\pi} \left[ {\Gamma_1\over 4} r_1 + m_1 f_1 \right]
- {\lambda^2_2\over\pi} \left[ {\Gamma_2\over 4} r_2 + m_2 f_2 \right],
\label{rhs2a}
\end{equation}
\begin{eqnarray}
{1\over\pi}\int^{s_0}_0 \hspace{-1mm} ds\,s\, \mbox{Im} F_2(s) =
  \lambda^2 m^3 
&+& {\lambda^2_1\over\pi} \left[ (s_0-m^2){\Gamma_1\over 2} 
+ m_1(m_1^2-3\Gamma_1^2/4) f_1 + {\Gamma_1\over 4} (3m_1^2-\Gamma_1^2/4) r_1 \right]
 \nonumber \\
&-& {\lambda^2_2\over\pi} \left[ (s_0-m^2){\Gamma_2\over 2} 
+ m_2(m_2^2-3\Gamma_2^2/4) f_2 + {\Gamma_2\over 4} (3m_2^2-\Gamma_2^2/4) r_2 \right],
\label{rhs2b}
\end{eqnarray}
\begin{eqnarray}
&& {1\over\pi}\int^{s_0}_0 \hspace{-1mm} ds\,s^2\, \mbox{Im} F_2(s) =
  \lambda^2 m^5 
 \nonumber \\
&+& {\lambda^2_1\over\pi} {\Gamma_1\over 2} 
 \left[ {s_0^2-m^4\over 2} + (s_0-m^2)(3m_1^2-{\Gamma_1^2\over 4}) 
+ {(2m_1^6-5m_1^4\Gamma_1^2+{5m_1^2\Gamma_1^4\over 8} \over m_1\Gamma_1} f_1
+ ({5m_1^4\over 2}-{5m_1^2\Gamma_1^2\over 4}+{\Gamma_1^4\over 32}) r_1 \right]
 \nonumber \\
&-& {\lambda^2_2\over\pi} {\Gamma_2\over 2} 
 \left[ {s_0^2-m^4\over 2} + (s_0-m^2)(3m_2^2-{\Gamma_2^2\over 4}) 
+ {(2m_2^6-5m_2^4\Gamma_2^2+{5m_2^2\Gamma_2^4\over 8} \over m_2\Gamma_2} f_2
+ ({5m_2^4\over 2}-{5m_2^2\Gamma_2^2\over 4}+{\Gamma_2^4\over 32}) r_2 \right],
\label{rhs2c}
\end{eqnarray}
where we have used the following definitions:
\begin{equation}
f_i \equiv \tan^{-1} \left( {s_0-m_i^2+\Gamma_i^2/4 \over m_i\Gamma_i} \right)
         + \tan^{-1} \left( {m_i^2-m^2-\Gamma_i^2/4 \over m_i\Gamma_i} \right), \hspace{1mm} (i=1,2),
\end{equation}
\begin{equation}
r_i \equiv \ln { (s_0-m_i^2+\Gamma_i^2/4)^2 + m_i^2\Gamma_i^2 \over 
  (m_i^2-m^2-\Gamma_i^2/4)^2 + m_i^2\Gamma_i^2 }, \hspace{1mm} (i=1,2).
\end{equation}
The above six expressions in Eq.~(\ref{rhs1a}) to Eq.~(\ref{rhs2c}) 
form the right-hand-side (RHS) of the sum rules.
Note that the couplings appear linearly, while the masses and widths highly non-linearly.

\end{widetext}

\section{Results and Discussions}

\begin{table*}[t]
\caption{The calculated six parameters ($\lambda_1^2$,$m_1$,$\Gamma_1$) and
($\lambda_2^2$, $m_2$, $\Gamma_2$) for the first two excited states in the baryon octet.
The ground-state pole ($\lambda^2$,$m$) is taken as input to the calculation.
The $c$'s are rescaled couplings $c_1=(2\pi)^4\lambda_1^2$ and so on.
The cutoff threshold $s_0$ is varied in each case to show sensitivity of the results to this parameter.
For each parameter, the range allowed within a certain global accuracy (last column)  
in the solutions is presented. For comparison purposes, 
the experimental values taken from the PDG~\protect\cite{pdg04} are also displayed.}
\label{tab1}

\begin{tabular}{c|cccccccccc}
\hline
 Baryon & $s_0$ & $c$ & $m$  & $m_1$ & $m_2$ & $\Gamma_1$ & $\Gamma_2$ & $c_1$ & $c_2$ & Accuracy  \\
& (GeV)  & (GeV$^6$) & (GeV) & (GeV) & (GeV) & (GeV) & (GeV) & (GeV$^6$) & (GeV$^6$) & (\%) \\ 
\hline
$N$
& 2.4 & 1.50 & 0.94 & 1.46-1.54 & 1.32-1.75 & 0.01-0.11 & 0.01-0.34 & 0.7-1.0 & 0.01-0.12 & $<6$ \\
& 2.5 & 1.50 & 0.94 & 1.46-1.57 & 1.32-1.78 & 0.01-0.10 & 0.01-0.35 & 0.8-1.2 & 0.01-0.12 & $<5.5$ \\
& 2.6 & 1.50 & 0.94 & 1.48-1.60 & 1.34-1.85 & 0.01-0.13 & 0.01-0.35 & 0.9-1.3 & 0.01-0.13 & $<4.5$ \\
Expt.
&     &      & 0.94 & 1.44 & 1.54 & 0.35 & 0.15 &           &         &      \\
\hline
$\Sigma$
& 3.0 & 2.50 & 1.19 & 1.72-1.73 & 1.74-2.02 & 0.01-0.02 & 0.01-0.34 & 0.80-1.25 & 0.1-0.9 & $<6$ \\
& 3.1 & 2.50 & 1.19 & 1.73-1.75 & 1.60-2.07 & 0.01-0.04 & 0.01-0.33 & 0.85-1.35 & 0.1-0.8 & $<5$  \\
& 3.2 & 2.50 & 1.19 & 1.72-1.78 & 1.62-2.08 & 0.01-0.07 & 0.01-0.33 & 0.95-1.40 & 0.1-1.1 & $<5$ \\
Expt.
&     &      & 1.19 & 1.66 & 1.75 & 0.10 & 0.09 &           &         &      \\
\hline
$\Lambda_o$
& 3.30 & 2.66 & 1.12 & 1.79-1.81 & 1.72-1.97 & 0.01-0.025 & 0.02-0.33 & 1.2-1.75& 0.15-0.8& $< 4$ \\
& 3.40 & 2.66 & 1.12 & 1.78-1.83 & 1.6-1.99  & 0.01-0.05  & 0.01-0.34 & 1.35-1.65 & 0.2-1.0& $< 4$ \\
& 3.50 & 2.66 & 1.12 & 1.79-1.86 & 1.6-1.98  & 0.01-0.06  & 0.01-0.34 & 1.45-1.8  & 0.25-0.9& $< 4$  \\
Expt.
&     &       & 1.12 & 1.60 & 1.67 & 0.15 & 0.035 &           &         &      \\
\hline
$\Xi$
& 4.00 & 3.56 & 1.32 & 1.97-1.99 & 1.69-2.04 & 0.01-0.03 & 0.01-0.34 & 1.9-2.6 & 0.6-1.7 & $< 2$ \\
& 4.15 & 3.56 & 1.32 & 1.97-2.03 & 1.74-2.07 & 0.01-0.07 & 0.01-0.34 & 2.1-3.2 & 0.7-1.5 & $< 2$ \\
& 4.30 & 3.56 & 1.32 & 1.98-2.06 & 1.77-2.10 & 0.01-0.06 & 0.01-0.34 & 2.5-3.0 & 0.9-1.4 & $< 2$ \\
Expt.
&     &       & 1.32 & 1.95 &      & 0.06 &      &           &         &      \\
\hline
\end{tabular}
\end{table*}
By equating the two sides of the moments of the spectral functions we get six sum rules. 
Mathematically, the problem boils down to finding the solutions to a 
system of six simultaneous, nonlinear equations of the form
\begin{equation}
\mbox{LHS}_i(s_0,\mbox{QCD})=
\mbox{RHS}_i(s0,\lambda^2,\lambda^2_1,\lambda^2_2,m,m_1,m_2,\Gamma_1,\Gamma_2),
\end{equation}
with $i=1,6$.
They can be used to solve for six unknowns, which we take as the two masses ($m_1,m_2$), 
two widths ($\Gamma_1,\Gamma_2$), and two couplings ($\lambda^2_1,\lambda^2_2$) for the 
two excited states. 
The remaining parameters are taken as input:
the ground-state pole ($m,\lambda^2$), as well as the continuum cutoff $s_0$.
The ground-state pole has been well-studied under the Borel sum rules 
using generalized interpolating fields~\cite{Lee02}.
For the cutoff $s_0$, we take values consistent with the observed spectrum~\cite{pdg04}, and vary 
it to see any sensitivity to this parameter.
For the QCD parameters in the LHS we use the following notations: 
$a=-(2\pi)^2\,\langle\bar{u}u\rangle$, 
$m^2_0=\langle\bar{q}g\sigma\cdot Gq\rangle/\langle\bar{q}q\rangle$,
$b=\langle g^2_c\, G^2\rangle$, 
$f_s=\langle\bar{s}s\rangle/\langle\bar{u}u\rangle$.
The four-quark condensate is parameterized using 
factorization approximation: $\langle\bar{u}u\bar{u}u\rangle=\kappa_v\langle\bar{u}u\rangle^2$, 
where the parameter $\kappa_v$ accounts for its violation. 
The anomalous dimension corrections of the various operators are taken into account 
via the factors $L^\gamma=\left[{\alpha_s(\mu^2)/ \alpha_s(s_0)}\right]^\gamma
=\left[{\ln(s_0/\Lambda_{QCD}^2)/
\ln(\mu^2/\Lambda_{QCD}^2)}\right]^\gamma$, where $\gamma$ is the appropriate anomalous 
dimension, $\mu$=500 MeV is the renormalization scale, and $\Lambda_{QCD}=0.15$ GeV is the QCD scale parameter.
The function $r(s_0)=\ln(s_0/\mu^2)-\gamma_{EM}$ with $\gamma_{EM}=0.577$, the Euler-Mascheroni constant. 
The numerical values, we use:
$a=0.52 \mbox{GeV}^3$, $b=1.2 \mbox{GeV}^4$, $m_0^2=0.8 \mbox{GeV}^2$, $\kappa_v=2$, 
$\bar{\alpha}_s/\pi=0.1$, $m_s=0.15$ GeV, $f_s=0.8$. 
In addition, we rescale the couplings to their more natural values by $c_1=(2\pi)^4\lambda_1^2$ 
and so on.
There is also the issue of optimal mixing parameter $t$ in the interpolating field.
We take $t=-0.8$ to get enhanced contributions of excited states, 
since $t=-1$ has been found not to couple strongly to the negative-parity state 
in the nucleon channel~\cite{Jido97}.
Since the couplings in~\cite{Lee02} were obtained for $t=-1$, 
we take the lowest values of the coupling in the given range.
The six equations are solved simultaneously by the Multidimensional secant Broyden's method 
from Numerical Recipes~\cite{nr92}. 
We also use the Globally-convergent Newton-Raphson method for checking purposes.
As a consistency check, we also solved the system of four equations by eliminating explicitly
$\lambda_1^2$ and $\lambda_2^2$ using the sum rules corresponding to Eq.~(\ref{rhs1a}) and Eq.~(\ref{rhs2a}).
All these methods give consistent results which are summarized in Table~\ref{tab1}.
 
As a first step, we revisited the nucleon channel using $t=-0.8$ rather than $t=-1$ as done 
in~\cite{Singh94}.  We have also taken $c=1.5\mbox{GeV}^6$, whereas in~\cite{Singh94}  
$c\approx 0.8\mbox{GeV}^6$ was used. Still our results come close to those obtained in~\cite{Singh94}.
We solved the equations by satisfying them within a global accuracy of few percent.
In this way, each parameter is allowed to vary in a certain range reflecting the stability of these 
parameters. We found it difficult to achieve better accuracies than the ones listed.
Overall, the results for the 1st excited state ($m_1,\Gamma_1$) are better than those for the 
2nd excited state ($m_2,\Gamma_2$). The $m_1$ is fairly stable. 
The $\Gamma_1$ is better constrained in the strange channels than in the nucleon channel.
The $c_1$ values, which are a measure of the ability of the interpolating fields to excite 
the state from the QCD vacuum,  are first-time information from the perspective of QCD. 
For the negative-parity states, whose contribution appears with the same signs as  
for positive-parity states in $F_1$ and opposite signs in $F_2$  sum-rules, 
the scatter in the numerical values obtained is larger, as noted earlier as well~\cite{Singh94}. 
We have restricted the values of $\Gamma_2$ from the upper side in each case to about 
0.34 GeV while the values of other parameters are as obtained without any restriction.
The results are fairly stable against variations in the cutoff parameter $s_0$.
The lowest $s_0$ in each channel corresponds to a mass value ($\sqrt{s_0}$) 
sightly above the value of $m_2$ obtained in that channel, which is quite reasonable,
and gives the lowest scatter in the values of physical parameters. 
In the $\Sigma$ and $\Lambda$ channels, $m_1$ is overestimated as compared to the experimental values. 
In the $\Xi$ channel, two spin-1/2 excited states are listed in PDG at 1690 MeV and 1950 MeV with 
3 stars and unassigned spin-parity. Our result favors the 1950 MeV state as the first excited state 
with positive-parity.

\section{Conclusion}
\label{con}

We have extended the study of nucleon excited states in~\cite{Singh94} 
to the strange members of the baryon octet, 
using the combination of Gauss-Weierstrass (GW) transform and finite energy sum rules (FESR).
Taking the ground-state pole as input, the masses and widths of the first two excited states 
of opposite parity can be computed in this approach, 
a possibility not afforded in the conventional Borel-based QCD sum rules. 
Furthermore, by changing the overall sign of the phenomenological side, 
the same approach can be used to study 
a particle channel that has the reverse parity ordering: ground state (negative), 1st excited (negative), 
2nd excited (positive).
A case in point in the physical spectrum is the flavor-singlet, spin-1/2 lambda sector 
where the ground state $\Lambda_S(1405)1/2-$ has negative parity.
A calculation is underway to understand this channel in the present approach.

Overall, our results show that the first excited state is well-probed by this method. 
The mass is stable and consistent with experiment, and the width has a relatively small uncertainty range.
On the other hand, the second excited state is less well-constrained; both the mass and width have 
larger scatter than those for the first excited state. But the second excited state can be well-described 
by a different method: the Borel-based parity-projected QCD sum rules~\cite{Jido97,Lee06}.
Taken together, the QCD sum rule method in its three variants (Borel-based conventional, 
Borel-based parity-projected, and GW-based FESR) gives access to the lowest three states 
in a given particle channel from a non-pertubative perspective of QCD.

\begin{acknowledgments}
This work is supported in part by the U.S. Department of Energy under grant DE-FG02-95ER40907.
JPS thanks the hospitality at GWU, where part of the work was done.
\end{acknowledgments}

\appendix
\begin{widetext}
\section{Wilson Coefficients}

Here we list the coefficients appearing in Eq.~(\ref{F1}) and Eq.~(\ref{F2}). 
Those which are common for all the members in the octet family are
(ignoring flavor breaking in radiative corrections)~\cite{Singh94,Lee02}:
%
\begin{equation}
A={1\over 32(2\pi)^4} (5+2t+5t^2) (1+{71\over 12}{\alpha_s\over \pi}),
\hspace{1mm}
B={-1\over 64(2\pi)^4} (5+2t+5t^2) ({\alpha_s\over \pi}),
\end{equation}
\begin{equation}
A_4={1\over 64(2\pi)^2} (5+2t+5t^2),
\hspace{1mm}
B_6={1\over 18} (62-4t-46t^2) ({\alpha_s\over \pi}).
\end{equation}
The other coefficients are member-dependent. For the $\Sigma$,
\begin{equation}
H_1={1\over 4(2\pi)^4} (1-t)^2,
\hspace{1mm}
C_3 = {-1\over 4(2\pi)^2} \left[ (6+f_s)(1+{15\over 14}{\alpha_s\over \pi})
        -2f_s t (1+{3\over 2}{\alpha_s\over \pi})
        -(6-f_s) t^2 (1+{7\over 10}{\alpha_s\over \pi}) \right],
\end{equation}
\begin{equation}
E_4 = {-1\over 8(2\pi)^2} \left[ (12-5f_s) -2f_s t -(12+5f_s) t^2 \right],
\hspace{1mm}
C_5 = {3\over 4(2\pi)^2} \left[ (1+{79\over 18}{\alpha_s\over \pi})
        - t^2 (1+{103\over 18}{\alpha_s\over \pi}) \right],
\end{equation}
\begin{equation}
D_5 = {3\over 4(2\pi)^2} (-{7\over 12}) (1-t^2) ({\alpha_s\over \pi}),
\hspace{1mm}
H_5 = {-1\over 32(2\pi)^2}  (1-t)^2,
\end{equation}
\begin{equation}
A_6 = {1\over 6} \left[ (6f_s+1)(1-{43\over 42}{\alpha_s\over \pi})
        -2 t (1-{1\over 6}{\alpha_s\over \pi})
        -(6f_s-1) t^2 (1-{29\over 30}{\alpha_s\over \pi}) \right],
\end{equation}
\begin{equation}
E_6 = {-1\over 24(2\pi)^2}
 \left[ (4f_s+21+18r(s_0)) + 4 f_s t + (4f_s-21-18r(s_0)) t^2 \right],
\end{equation}
\begin{equation}
C_7 = {-1\over 288} \left[ (24-5f_s) + 10 f_s t - (24+5f_s) t^2 \right],
\hspace{1mm}
H_7 = {1\over 6} \left[ (5-3f_s) + 2 t + (5+3f_s) t^2 \right],
\end{equation}
\begin{equation}
A_8 = {-1\over 24} \left[ (1+12f_s) - 2 t - (12f_s-1) t^2 \right].
\end{equation}
%
For the octet $\Lambda_O$,
\begin{equation}
H_1={11+2t-13t^2\over 12(2\pi)^4},
\hspace{1mm}
C_3 = {-1\over 12(2\pi)^2} \left[ (10+11f_s)(1+{15\over 14}{\alpha_s\over \pi})
       +(-8+2f_s) t (1+{3\over 2}{\alpha_s\over \pi})
        -(2+13f_s) t^2 (1+{7\over 10}{\alpha_s\over \pi}) \right],
\end{equation}
\begin{equation}
E_4 = {-1\over 24(2\pi)^2} \left[ (20-15f_s) - (16+6f_s) t - (4+15f_s) t^2 \right],
\hspace{1mm}
C_5 = {1+2f_s\over 4(2\pi)^2} \left[ (1+{79\over 18}{\alpha_s\over \pi})
        - t^2 (1+{103\over 18}{\alpha_s\over \pi}) \right],
\end{equation}
\begin{equation}
D_5 = {1+2f_s\over 4(2\pi)^2} (-{7\over 12}) (1-t^2) ({\alpha_s\over \pi}),
\hspace{1mm}
H_5 = {1\over 96(2\pi)^2}  (13-2t-11t^2),
\end{equation}
\begin{equation}
A_6 = {1\over 18} \left[ (10f_s+11)(1-{43\over 42}{\alpha_s\over \pi})
        - (2-8f_s) t (1-{1\over 6}{\alpha_s\over \pi})
        - (2f_s+13) t^2 (1-{29\over 30}{\alpha_s\over \pi}) \right],
\end{equation}
\begin{equation}
E_6 = {1\over 24(2\pi)^2}
 \left[ (4f_s-5-6r(s_0)) + (4+4f_s) t + (4f_s+1+6r(s_0)) t^2 \right],
\end{equation}
\begin{equation}
C_7 = {-1\over 864} \left[ (4+53f_s) + (40-10f_s) t - (44+33f_s) t^2 \right],
\hspace{1mm}
H_7 = {1\over 18} \left[ (15-5f_s) + (6+4f_s) t + (15+f_s) t^2 \right],
\end{equation}
\begin{equation}
A_8 = {-1\over 72} \left[ (23+16f_s) + (2-8f_s) t - (25+8f_s) t^2 \right].
\end{equation}
For the $\Xi$,
\begin{equation}
H_1={3\over 2(2\pi)^4} (1-t^2),
\hspace{1mm}
C_3 = {-1\over 4(2\pi)^2} \left[ (6f_s+1)(1+{15\over 14}{\alpha_s\over \pi})
        -2 t (1+{3\over 2}{\alpha_s\over \pi})
        -(6f_s-1) t^2 (1+{7\over 10}{\alpha_s\over \pi}) \right],
\end{equation}
\begin{equation}
E_4 = {-3\over 4(2\pi)^2} \left[ (2-f_s) -2f_s t - (2+f_s) t^2 \right],
\hspace{1mm}
C_5 = {3f_s\over 4(2\pi)^2} \left[ (1+{79\over 18}{\alpha_s\over \pi})
        - t^2 (1+{103\over 18}{\alpha_s\over \pi}) \right],
\end{equation}
\begin{equation}
D_5 = {3f_s\over 4(2\pi)^2} (-{7\over 12}) (1-t^2) ({\alpha_s\over \pi}),
\hspace{1mm}
H_5 = {3\over 16(2\pi)^2}  (1-t^2),
\end{equation}
\begin{equation}
A_6 = {f_s\over 6} \left[ (f_s+6)(1-{43\over 42}{\alpha_s\over \pi})
        -2f_s t (1-{1\over 6}{\alpha_s\over \pi})
        +(f_s-6) t^2 (1-{29\over 30}{\alpha_s\over \pi}) \right],
\end{equation}
\begin{equation}
E_6 = {-1\over 24(2\pi)^2}
 \left[ (15-f_s+18r(s_0)) - 10f_s t - (15+f_s+18r(s_0)) t^2 \right],
\end{equation}
\begin{equation}
C_7 = {-1\over 288} \left[ (24f_s-5) + 10 t - (24f_s+5) t^2 \right],
\hspace{1mm}
H_7 = {f_s\over 2} \left[ (3-f_s) + 2 t + (3+f_s) t^2 \right],
\end{equation}
\begin{equation}
A_8 = {-f_s\over 24} \left[ (f_s+12) - 2f_s t - (f_s-12) t^2 \right].
\end{equation}

\end{widetext}


\end{document}